\documentclass[11pt]{article}
\usepackage{moriond,epsfig}

\def\Journal#1#2#3#4{{#1} {\bf #2}  (#3) #4}

\def\NIM{\em Nucl. Instrum. Methods}

\def\NPB{{\em Nucl. Phys.} }
\def\PLB{{\em Phys. Lett.} }
\def\PRL{\em Phys. Rev. Lett.}

\def\PRD{{\em Phys. Rev.} }

\def\JHEP{{\em J. High Energy Phys.}}
\def\EPJ{{\em Eur Phys. J}}
\def\RMP{{\em Rev. Mod. Phys}}
\def\JPGH{{\em J. Phys.}}
\def\arXivid#1{{\ttfamily {arXiv:#1}}}

\def\ko{K^0}
\def\kb{\bar{K^0}}

\def\ab{\bar{\alpha}}
\def\be{\begin{equation}}
\def\ee{\end{equation}}
\def\bea{\begin{eqnarray}}
\def\eea{\end{eqnarray}}

\def\ifm#1{\relax\ifmmode#1\else$#1$\fi}
 \def\up#1{\ifm{^{#1}}} 
\def\plm{\ifm{\,\pm}\,}   \def\ab{\ifm{\sim}}      
\def\DAF{DA\char8NE}  \def\sig{\ifm{\sigma}}
\def\f{\ifm{\phi}}  
\def\epm{\ifm{e^+e^-}}
 \def\kl{\ifm{K_L}}
\def\ks{\ifm{K_S}}  \def\kpm{\ifm{K^\pm}}  \def\kp{\ifm{K^+}}  \def\km{\ifm{K^-}}
   \def\ko{\ifm{K^0}}      \def\epm{\ifm{e^+e^-}}
\def\kb{\ifm{\rlap{\kern.3em\raise1.9ex\hbox to.6em{\hrulefill}} K}}
  \def\sig{\ifm{\sigma}}  

  \def\po{\ifm{\pi^0}}    \def\rmk{\rm\kern.5mm }

\def\figb#1;#2;{\parbox{#2cm}{\epsfig{file=#1.eps,width=#2cm}}}
\def\figbc#1;#2;{\cl{\figb #1;#2;}}     
\def\ie{{\it\kern-1pt i.\kern-.5pt e.\kern-.2pt}}  
\let\cl=\centerline
\newcommand{\Ppim}{\ensuremath{\pi^-}}
\newcommand{\Ppin}{\ensuremath{\pi^0}}
\newcommand{\Ppip}{\ensuremath{\pi^+}}
\newcommand{\Vus}{\ensuremath{|V_{us}|}}
\newcommand{\Vud}{\ensuremath{|V_{ud}|}}

\newcommand{\fVus}{\ensuremath{|f_+(0)\,V_{us}|}}

\newcommand{\fo}{\ensuremath{f_+(0)}}
\newcommand{\fkfp}{\ensuremath{f_K/f_\pi}}
\newcommand{\dSU}{\ensuremath{\delta^{\rm SU(2)}}}
\newcommand{\dEM}{\ensuremath{\delta^{\rm EM\vphantom{()}}}}

\newcommand{\fphat}{\mbox{$\tilde{f}_+(t)$}}

\newcommand{\fzhat}{\mbox{$\tilde{f}_0(t)$}}

\newcommand{\lam}{\mbox{$\lambda_+$}}
\newcommand{\lamp}{\mbox{$\lambda'_+$}}
\newcommand{\lampp}{\mbox{$\lambda''_+$}}
\newcommand{\lamo}{\mbox{$\lambda_0$}}
\newcommand{\lamop}{\mbox{$\lambda'_0$}}
\newcommand{\lamopp}{\mbox{$\lambda''_0$}}

\catcode`@=11 
\newdimen\z@ \z@=0pt 
\newskip\z@skip \z@skip=0pt plus0pt minus0pt
\def\m@th{\mathsurround=\z@}
\def\ialign{\everycr{}\tabskip\z@skip\halign} 
\def\eqalign#1{\null\,\vcenter{\openup\jot\m@th
  \ialign{\strut\hfil$\displaystyle{##}$&$\displaystyle{{}##}$\hfil
      \crcr#1\crcr}}\,}
\catcode`@=12 

\newcommand{\BR}[1]{\ensuremath{{\rm BR}(#1)}}

\renewcommand{\Re}[1]{\ensuremath{{\rm Re}\:#1}}
\renewcommand{\Im}[1]{\ensuremath{{\rm Im}\:#1}}
\newcommand{\abs}[1]{\ensuremath{\left|#1\right|}}
\newcommand{\VSS}[3]{\ensuremath{#1\pm#2_{\rm stat}\pm#3_{\rm syst}}}

\def\ord#1;{\ifm{{\mathcal O}(#1)}}

\newcommand{\Eq}[1]{eq. \ref{#1}}

\newcommand{\Fig}[1]{figure \ref{#1}}
\newcommand{\Ref}[1]{ref. \cite{#1}}

\begin{document}
\vspace*{4cm}
\title{Recent results from KLOE}

\author{Marianna Testa for the KLOE Collaboration\footnote{
F.~Ambrosino,
A.~Antonelli,
M.~Antonelli,
F.~Archilli,
C.~Bacci,
P.~Beltrame,
G.~Bencivenni,
S.~Bertolucci,
C.~Bini,
C.~Bloise,
S.~Bocchetta,
F.~Bossi,
P.~Branchini,
R.~Caloi,
P.~Campana
G.~Capon,
T.~Capussela,
F.~Ceradini,
F.~Cesario,
S.~Chi,
G.~Chiefari,
P.~Ciambrone,
F.~Crucianelli,
E.~De~Lucia,
A.~De~Santis,
P.~De~Simone,
G.~De~Zorzi,
A.~Denig,
A.~Di~Domenico,
C.~Di~Donato,
B.~Di~Micco,
A.~Doria,
M.~Dreucci,
G.~Felici,
A.~Ferrari,
M.~L.~Ferrer,
S.~Fiore,
C.~Forti,
P.~Franzini,
C.~Gatti,
P.~Gauzzi,
S.~Giovannella,
E.~Gorini,
E.~Graziani,
W.~Kluge,
V.~Kulikov,
F.~Lacava,
G.~Lanfranchi,
J.~Lee-Franzini,
D.~Leone,
M.~Martemianov,
M.~Martini,
P.~Massarotti,
W.~Mei,
S.~Meola,
S.~Miscetti,
M.~Moulson,
S.~M\"uller,
F.~Murtas,
M.~Napolitano,
F.~Nguyen,
M.~Palutan,
E.~Pasqualucci,
A.~Passeri,
V.~Patera,
F.~Perfetto,
M.~Primavera,
P.~Santangelo,
G.~Saracino,
B.~Sciascia,
A.~Sciubba,
A.~Sibidanov,
T.~Spadaro,
M.~Testa,
L.~Tortora,
P.~Valente,
G.~Venanzoni,
R.~Versaci,
G.~Xu.}}

\address{INFN-LNF, Via E. Fermi 40,\\
I-00044 Frascati, Italy}

\maketitle\abstracts{
In this report  I will present  the recent results on $K$ mesons from the KLOE experiment  at the DAFNE $e^+e^-$ collider working at the center of mass energy $\sim$ 1GeV $\sim m_{\phi}$. They  include $V_{us}$ determinations, the test on the  unitarity of the first row of the CKM matrix  and the related experimental measurements. Tests of  lepton universality from leptonic  and semileptonic decays will be also discussed. Then I will present tests of quantum coherence, $CPT$ and Lorentz symmetry    performed  by studying the time evolution of the neutral kaon system. }

\section{The KLOE experiment}
The KLOE detector operates at \DAF, an $\epm$ collider working at the center of mass energy $W\sim m_{\f} \sim 1.02$ GeV.   The \f\ mesons are produced essentially at rest and decay to $\ks\kl$ ($K^+K^-$) $\sim$ 34\% ($\sim$ 49\%) of the times. The $K$ mesons are produced in a pure  $J^{PC}=1^{--}$ coherent quantum state, so that observation of a \ks\ ($K^+$) in an event signals (tags) the presence of a \kl\ ($K^-$) and vice-versa: highly pure, almost monochromatic, back-to-back \ks\ ($K^+$) and \kl\ ($K^-$) beams  can be obtained.  Moreover \ks\ and \kl\ are distinguishable on the basis of their decay length: $\lambda_S \sim 0.6$ cm and $\lambda_L \sim 340 $ cm. \\
The KLOE detector consists essentially of a drift chamber (DC), surrounded by an electromagnetic calorimeter (EMC). The DC~\cite{KLOE+02:DC} is a cylinder of 4 m diameter and 3.3 m in length which constitutes a large fiducial volume for  \kl\ decays ($\sim$1/2 of $\lambda_L$). The momentum resolution for tracks at large polar angle is $\sigma_p/p \le 0.4\%$. The EMC~\cite{KLOE+02:EmC} is a lead-scintillating fiber calorimeter consisting of a barrel and two endcaps, which cover 98\% of the solid angle. The energy resolution is $\sigma_E/E \sim 5.7\%/\sqrt{E(\mbox{GeV})}$. The intrinsic time resolution is $\sigma_T = 54 \mbox{ps}/\sqrt{E(\mbox{GeV})} \oplus 50 \mbox{ps}$. A superconducting coil surrounding the barrel provides a 0.52 T magnetic field.\\
The present report is  based on a first data sample of $\sim$500 pb$^{-1}$, except for quantum coherence, $CPT$ and Lorentz symmetry tests; at present KLOE has about 2.2 fb$^{-1}$ on disk.

\section{ $V_{us}$ determination}
In the Standard Model, the coupling of the $W$ boson to the weak charged current is written as
\begin{equation}
{g\over\sqrt2}W_\alpha^+(\rlap{\bf U}\raise2.2ex\hbox to.6em{\hrulefill}\kern.3em_L\,{\bf V}_{\rm CKM}\gamma^\alpha\,{\bf D}_L+\bar e_L\gamma^\alpha\nu_{e\,L}+\bar \mu_L\gamma^\alpha\nu_{\mu\,L}+\bar \tau_L\gamma^\alpha\nu_{\tau\,L})\ +\ \mbox{h.c.},
\end{equation}
where ${\bf U}^{\rm T}=(u,c,t)$, ${\bf D}^{\rm T}=(d,s,b)$ and $L$ is for lefthanded.
In the coupling above there is only one coupling constant for leptons and quarks. Quarks are mixed by the Cabibbo-Kobayashi-Maskawa matrix, ${\bf V}_{\rm CKM}$, which must be unitary. \\
The most precise check on the unitarity of the  ${\bf V}_{\rm CKM}$ matrix is provided by measurements of $|V_{us}|$ and $|V_{ud}|$, the contribution of $V_{ub}$ being at the level of 10\up{-5}.
\Vus\ may be extracted by the measurements of the semileptonic decay rates, fully inclusive of radiation, which are given by: 
\begin{equation}
\Gamma(K_{\ell3(\gamma)}) =
\frac{C_K^2 G_F^2 M_K^5}{192\pi^3}\,S_{\rm EW}\,\Vus^2\,\abs{\fo}^2\,
I_{K\ell}\,\left(1 + \dSU_K + \dEM_{K\ell}\right)^2.
\label{eq:Vus}
\end{equation}
In the above expression, the index $K$ denotes $\ko\to\pi^\pm$ and $\kpm\to\po$ transitions, for which $C_K^2 =1$ and 1/2, respectively. $M_K$ is the appropriate kaon mass, $S_{\rm EW}$ is the universal short-distance electroweak correction \cite{Sir82:SEW} and $\ell=e,\:\mu$. Following a common convention, $\fo \equiv f_+^{K^0\pi^-}(0)$. The mode dependence is contained in the $\delta$ terms: the long-distance electromagnetic (EM) corrections, which depend on the meson charges and lepton masses and the SU(2)-breaking corrections, which depend on the kaon species \cite{C+02:Kl3rad}.
$I_{K\ell}$ is the integral of the dimensionless Dalitz-plot density  over the physical region for non radiative decays and includes $|\tilde f_{+,\,0}(t)|^2$, the reduced form factor, defined below.

\Vus\ can be also extracted from $K\rightarrow\mu\nu$ decays using the relation
\begin{equation}
{\Gamma(K_{\mu2(\gamma)})\over\Gamma(\pi_{\mu2(\gamma)})}=%
{\Vus^2\over\Vud^2}\;{f_K^2\over f_\pi^2}\;%
{m_K\left(1-m^2_\mu/m^2_K\right)^2\over m_\pi\left(1-m^2_\mu/m^2_\pi\right)^2}\times(0.9930\pm0.0035),
\label{eq:fkfp}
\end{equation}
where  $f_\pi$ and $f_K$  are the pion- and kaon-decay constants and  the uncertainty in the numerical factor is dominantly from structure-dependent radiative corrections.  This ratio can be combined with direct measurements of \Vud\ to obtain \Vus. \\
 The measurement of $V_{us}$ from leptonic and semileptonic kaon decays allows both the test the unitarity of the CKM matrix and and the leptonic quark universality. 
  Moreover the universality of electron and muon interactions can be tested by measuring the ratio $\Gamma(K\to\pi \mu \nu)/\Gamma(K\to\pi e \nu)$ and the comparison between the measurement of $V_{us}$ from leptonic decays and that from semileptonic decays allows to put bounds on new physics.

The experimental inputs to  \Eq{eq:Vus} and \ref{eq:fkfp} are the semileptonic and leptonic  decay rates, fully inclusive of radiation,  \ie\ branching ratios (BR) and lifetimes, and the reduced form factors \fphat\ and \fzhat, whose behaviour as a function of $t$, the 4-momentum transfer squared $(P_K-p_{\pi})^2$,  is obtained from the decay pion spectra. Details on the  measurements and  the treatment of correlations can be found in \Ref{KLOE:Vus}.
In this report I will present  the recent measurement of the $K_{\mu3}$ form factors, the charged kaon life time, the BR($K^{\pm}_{l3}$) and  the  BR($K^{+}\to \Ppip\Ppin$)

\section{K$\mu$3 from factors}

The largest uncertainty in calculating \Vus\ from the decays rates is due to the difficulties in computing the matrix element $\langle \pi|J^{had}_\alpha|K\rangle$  which has the form:
\begin{equation} \langle \pi|J^{had}_\alpha|K\rangle = f_+(0)\times \left( (P+p)_{\alpha} f_+(t) + (P-p)_{\alpha} (f_0(t)- f_+(t) \Delta_{K\pi}/t \right)\end{equation} where $P(p)$ is the $K(\pi)$ momentum, $t = (P-p)^2$ and $\Delta_{K\pi} = M_K^2-m_{\pi}^2$.  The above equation defines the vector and scalar form factors (FF)  $f_+(t)= f_+(0)\tilde f_+(t)$ and $f_0(t)= f_+(0)\tilde f_0(t)$, which take into account the non point-like structure of the pions and kaons. The term $f_+(0)$ has been factored out, since the FFs must have the same value at $t=0$. 
If the FFs are expanded in powers of $t$ up to $t^2$ as
$\tilde{f}_{+,0}(t) = 1 + \lambda'_{+,0}~\frac{\displaystyle t}{\displaystyle m^2} +
  \frac{\displaystyle 1}{\displaystyle 2}\;\lambda''_{+,0}\,\left(\frac{\displaystyle t}{\displaystyle m^2}\right)^2,$
four parameters (\lamp, \lampp, \lamop\ and \lamopp) need to be
determined from the decay pion spectrum in order to be able to compute the
phase-space integral.
However, this parametrization of the form factors is problematic,
because the values for the $\lambda$s obtained from fits to the
experimental decay spectrum
are strongly correlated ~\cite{Fra07:Kaon}.
It is therefore   necessary to obtain a form for \fzhat\ and \fphat\ with at least $t$ and $t^2$ terms but with only one parameter.
The Callan-Treiman relation \cite{ct} fixes the
value of scalar FF at $t=\Delta_{K\pi}$ (the so-called Callan-Treiman point)
to the ratio of the pseudoscalar decay constants $f_K/f_\pi$.
$\tilde{f}_0(\Delta_{K\pi})=\frac{\displaystyle f_K}{\displaystyle f_\pi}\:{1\over \displaystyle f_+(0)}+\Delta_{\rm CT}$,
where $\Delta_{\rm CT}$, SU(2)-breaking correction \cite{dct}, is of {$\mathcal O$}($10^{-3})$. A recent dispersive parametrization
for the scalar form factor \cite{stern}, $\fzhat =\exp\left[\frac{\displaystyle t}{\displaystyle \Delta_{K\pi}} (\ln C - G(t))\right]$, allows the constraint given by the
Callan-Treiman relation to be exploited,  
such that $C=\tilde{f}_0(\Delta_{K\pi})$ and $\tilde{f}_0(0) = 1$.
 $G(t)$ is derived from $K\pi$ scattering data.
As suggested in \Ref{stern}, a good approximation to the dispersive parametrization  is
$\fzhat = 1 + \lamo{\displaystyle t\over \displaystyle m^2} + \frac{\displaystyle \lamo^2 + p_2}{\displaystyle 2}\left({\displaystyle t\over \displaystyle  m^2}\right)^2 + \frac{\displaystyle \lamo^3 + 3p_2\lamo + p_3}{\displaystyle 6}\left({\displaystyle t\over\displaystyle  m^2}\right)^3$
with $p_2$ and $p_3$ given in~\Ref{stern}.
Also for the vector FF we  make use of a  dispersive  parameterization~\cite{sternV}, twice substracted at $t=0$, $\fphat =\exp\left[\frac{\displaystyle t}{\displaystyle m^2_{\pi}} (\Lambda_+ + H(t))\right]$, where $H(t)$ is obtained from $K\pi$ scattering data and $\Lambda_+$ has to be determined from  the fit to experimental data.
\begin{figure}[h] 
\begin{center}
\includegraphics[width=7.0cm]{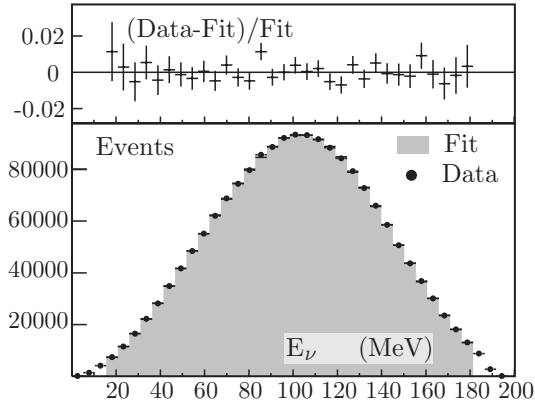}
\caption{Residuals of the fit (top plots) and $E_\nu$ distribution for data events superimposed on the fit result (bottom plot)\label{fig:km3}}
\end{center}
\end{figure}
At KLOE energies clean and efficient $\pi/\mu$ separation, required to measure the $t$ spectrum, is
 difficult. The FF parameters have   been therefore obtained from fits to the
distribution of the neutrino energy $E_\nu$ after integration over the pion energy. About 1.8 Million of $K_{\mu3}$   are selected by means of kinematic cuts, time of flight (TOF) measurements and calorimetric information. Details on the analysis can be found in \Ref{kmu3ff}.  
Using the dispersive  parameterizations for the vector and scalar FF's  and combing the $K_{\mu3}$  and $K_{e3}$  data,  we find 
 $   \lam  = ({25.7}\pm {0.4}\pm{0.4}\pm 0.2_{\rm {param}}) \times10^{-3}$ and 
$    \lamo = ({14.0}\pm {1.6}\pm {1.3}\pm 0.2_{\rm {param}}) \times10^{-3}$
with $\chi^2/{\rm dof}=2.6/3$ and a correlation coefficient of
$-0.26$. The result of the fit on $K_{\mu3}$ data is shown in \Fig{fig:km3}.
Preliminary results based on ~1$fb^{-1}$ have been also obtained and averaged with that presented above:
$ \lam  = ({26.0}\pm{0.5_{stat+syst}}) \times10^{-3}$
and $\lamo  = ({15.1}\pm{1.4}_{stat+syst}) \times10^{-3}$
\section {$\tau(K^{\pm})$, BR($K^{\pm}_{l3}$) and   BR($K^{+}\to \Ppip\Ppin$)}
We have combined the recent published  measurements of the semileptonic BRs and the charged kaon lifetime 
 to use them in the  evaluation of \Vus.

At KLOE, two methods are used to reconstruct the proper decay time distribution for charged kaons.
The first is to obtain the decay time from the
kaon path length in the DC, accounting for the continuous change
in the kaon velocity due to ionization energy losses.
A fit to the proper-time distribution in the interval from 15--35~ns
($1.6\tau_\pm$) gives the result $\tau_\pm = \VSS{12.364}{0.031}{0.031}$~ns.
Alternately, the decay time can be obtained from the precise measurement
of the arrival times of the photons from $\kp\to\pi^+\po$ decays.
In this case, a fit to the proper-time distribution in the interval from
13--42~ns
($2.3\tau_\pm$) gives the result $\tau_\pm = \VSS{12.337}{0.030}{0.020}$~ns.
Taking into account the statistical correlation between
these two measurements ($\rho=0.307$), we obtain
the average value $\tau_\pm = 12.347\pm0.030$~ns, see \cite{massar}.

\begin{figure}[h]
\begin{center}
\includegraphics[width=7.0cm]{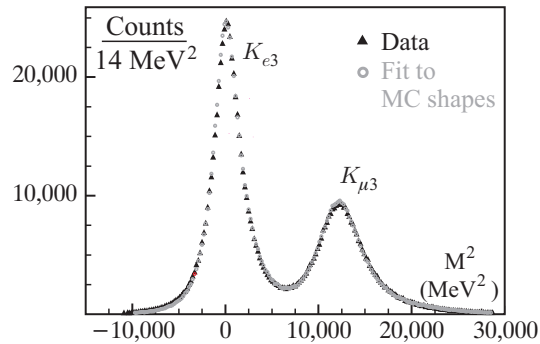}
\caption{Distribution of $m^2_l$, from TOF information, for $K^\pm_{l3}$ events.\label{fig:kl3ch}}
\end{center}
\end{figure}

To measure \BR{K^\pm_{e3}} and \BR{K^\pm_{\mu3}}, we use
both $K\to\mu\nu$ and $K\to\pi\po$ decays as tags.
We measure the semileptonic BRs separately for
\kp\ and \km. Therefore, \BR{K_{e3}} and \BR{K_{\mu3}}
are each determined from four independent
measurements (\kp\ and \km\ decays; $\mu\nu$ and $\pi\po$ tags).
Two-body decays are removed  by kinematics 
and  the photons from the \po\ are reconstructed to reconstruct
the \kpm\ decay point. From the TOF and momentum measurement for the lepton tracks, we obtain the $m^2_l$ distribution shown in \Fig{fig:kl3ch}. Further details are given in~\cite{barb}.
Using the above result for $\tau_\pm$ to estimate the fiducial volume acceptance, we obtain 
BR$(K_{e3})=0.04972\pm 0.00053$ and BR$(K_{\mu3})=0.03273\pm 0.00039$,
which we use in our evaluation of \Vus.

We have also obtained  a preliminary result on the BR($K^+ \to \Ppip\Ppin$), which  is crucial to perform the fit of all  \kpm\  BRs  and  for the \Vus\ determination  of several experiments (NA48, ISTRA+, E865) in the normalization  of the BRs ($K^{\pm}_{l3}$). About 800000 $K^+ \to \Ppip\Ppin$ have been select with kinematic cuts. Our preliminary result, BR($K^+ \to \Ppip\Ppin$) =  $(20.658 \pm 0.065 \pm 0.090)\%$, is lower than the PDG  value~\cite{PDG07} of about 1.3\%. Further details can be found in \Ref{brk+pp}.

\section{$|f_+(0)V_{us}|$ and  lepton universality} 
Using the BR($K^{0,\pm}_{l3}$), $\tau(\kl)$,  $\tau(\kpm)$ and the FFs  from the KLOE results and  $\tau(\ks)$ from the PDG~\cite{PDG07}, 
the values of  $|f_+(0)V_{us}|$  has been evaluated for $K_{Le3}$,  $K_{L\mu3}$, $K_{Se3}$, $K^{\pm}_{e3}$ and   $K^{\pm}_{\mu3}$ decay modes. 
The inputs from theory, according to \Eq{eq:Vus}, are the SU(2)-breaking correction evaluated with ChPT to \ord p^4;, as described
in \cite{su2breaking}, the long distance EM corrections to the full inclusive decay rate  evaluated with ChPT to
\ord e^2p^2; \cite{su2breaking} using low-energy constants
from ref. \cite{moussallam}.\\
The average on the five different determination obtained  taking into account all correlations is: $ |f_+(0)V_{us}| = 0.2157 \pm 0.0006$ with $\chi^2/{\rm dof}=7.0/4$.\\
Comparison of the values of \fVus\ for $K_{e3}$ and $K_{\mu3 }$ modes
provides a test of lepton universality. We calculate the following quantity
\begin{equation}
r_{\mu e} \equiv \frac{\fVus^2_{\mu3, \ {\rm exp}}}
                      {\fVus^2_{e3, \ {\rm exp}}} =
                 \frac{\Gamma_{\mu 3}}{\Gamma_{e3}}\:
                 {I_{e3}\left(1+\delta_{Ke}\right)^2\over
                  I_{\mu3}\left(1+\delta_{K\mu}\right)^2},
\label{eq:leptuniv}
\end{equation} where $\delta_{K\ell}$ stands for
$\dSU_K + \dEM_{K\ell}$. In the SM $r_{\mu e}=1$. Averaging between charged and neutral modes, we find $r_{\mu e} =  1.000\pm0.008.$  The sensitivity of this result is competitive with that obtained for $\pi\to l\nu$  and   $\tau \to l\nu$ decays~\cite{piuniv,DHZ06} whose accuracy is  $\sim 0.4\%$.

\section{Test of CKM unitarity}
\begin{figure}[h]
\begin{center}
\includegraphics[width=7.0cm]{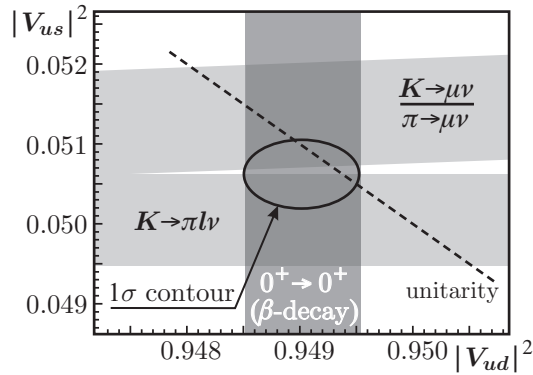}
\caption{KLOE results for $|V_{us}|^2$, $|V_{us}/V_{ud}|^2$ and $|V_{ud}|^2$ from $\beta$-decay measurements, shown as 2\sig\ wide grey bands. The ellipse is the 1 \sig\ contour from the fit. The unitarity constraint is illustrated by the dashed line.}\label{fig:vusvud}

\end{center}
\end{figure}

 To get the value of \Vus\ we have used the  recent determination of $\fo= 0.9644\pm0.0049$ from RBC and UKQCD Collaborations obtained  from a lattice calculation with $2+1$ flavors of dynamical domain-wall fermions \cite{rbcukqcd07:f0}. Using their value for \fo, our $K_{l3}$ results give $\Vus = 0.2237\plm 0.0013$. 
Additional information is provided by  the determination of the ratio $|V_{us}/V_{ud}|$, using \Eq{eq:fkfp}. From our measurements of BR($K_{\mu 2}$) and $\tau_{\pm}$, $\Gamma(\pi_{\mu2})$ from \Ref{PDG07} and the recent lattice determination of \fkfp\ from the HP\-QCD/UKQCD collaboration, \fkfp =1.189\plm0.007 \cite{hpqcdukqcd07:fkfp}, we obtain $|V_{us}/V_{ud}|^2$=0.0541 \plm0.0007. 
We perform a fit to the above ratio and our result $\Vus^2$=0.05002\plm0.00057 together with the result \Vud\up2\kern1mm=\kern1mm0.9490\plm 0.0005 from superallowed $\beta$-decays \cite{vud07ht}.
We find  $1-\Vus^2-\Vud^2=0.0004\pm0.0007\quad(\ab0.6\sigma)$ and 
confirm the unitarity of the CKM quark mixing matrix as applied to the first row. The result of the fit is shown in \Fig{fig:vusvud}.

\section{Bounds on new physics from $K_{l2}$ decays}
\begin{figure}[h]
\begin{center}
\includegraphics[width=7.0cm]{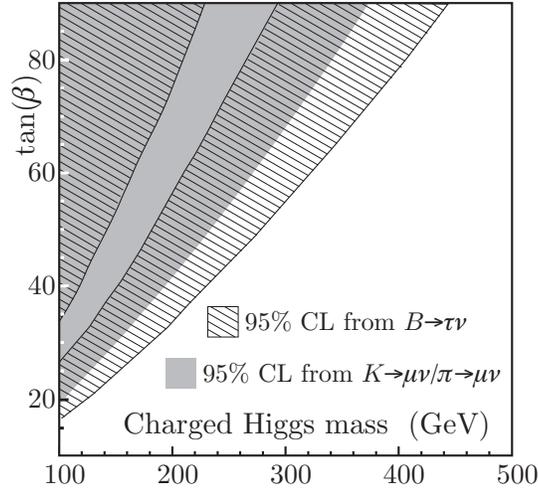}
\caption{Region in the $m_{H^+}$-$\tan\beta$ plane excluded by our result
for $R_{\ell23}$; the region excluded by measurements of
$\BR{B \rightarrow \tau \nu}$ is also shown.\label{fig:higgs}}
\end{center}
\end{figure}

 The comparison between the values for \Vus\ obtained from helicity-suppressed $K_{\ell2}$
decays and helicity-allowed $K_{\ell3}$ decays allows to  put bounds on new physics. We study the quantity $ R_{\ell23} = \left|  \frac{ \displaystyle V_{us}(K_{\mu2})}{\displaystyle  V_{us}(K_{\ell3})}\times
  \frac{\displaystyle  V_{ud}(0^+\to0^+)}{\displaystyle  V_{ud}(\pi_{\mu2})}  \right|$, which is unity in the SM, but would be affected only  in $V_{us}(K_{\mu2})$  by the presence of non-vanishing scalar or right-handed currents.
A scalar current due
to a charged Higgs exchange is expected to lower the value of
$R_{\ell23}$, which becomes (see \cite{IP06}): $R_{\ell23} = \left|   1 - \frac{\displaystyle  m^2_{K^+}}{\displaystyle  m^2_{H^+}}\:         \left(1 - \frac{\displaystyle  m^2_{\pi^+}}{\displaystyle  m^2_{K^+}}\right)\:     \frac{\displaystyle  \tan^2 \beta}{\displaystyle  1 + \epsilon_0\,\tan \beta} \right|$ with $\tan \beta$ the ratio of the two Higgs vacuum expectation values
in the MSSM and $\epsilon_0 \approx 0.01$ \cite{IR01}.
Using our result  on
$K_{\mu2}$ and $K_{\ell3}$ decays, the lattice
determinations of \fo\ and \fkfp\ and the value of \Vud\
discussed above, we obtain  $ R_{\ell23} = 1.008 \pm 0.008$. Fig.~\ref{fig:higgs} shows the region in the
$\{m_{H^+},\ \tan\beta\}$ plane excluded at 95\% CL by our result for $R_{\ell23}$.

The ratio $R_K=\frac{\displaystyle BR(K_{e2})}{\displaystyle BR(K_{\mu2})}$ is extremely well known in the SM, being almost free on hadronic uncertainties.  
Since the electron channel is helicity suppressed $R_K$ is sensitive to contributions from physics beyond the SM. Deviations up to few percent  on $R_K$ are expected in  minimal supersymmetric extensions of the SM and should be dominated by lepton-flavour violating contributions  with tauonic neutrinos emitted~\cite{masiero}. KLOE has selected about 8000 $K_{e2}$ events on 1.7 $pb^{-1}$ by performing a direct search without the tag of the other kaon. Background from $K_{\mu2}$ has been reduced by means of kinematic cuts and calorimeter particle identification. Our preliminary result, $R_K=(2.55\pm0.05\pm0.5)\times10^{-5}$, allows to put bounds on the charged Higgs mass and $\tan\beta$ for different slepton mass matrix off-diagonal elements $\Delta_{1,3}$. An accuracy of $\sim 1\%$ is expected increasing the  data sample analized, the control sample and Monte Carlo statistics.

\section{Test of quantum coherence, CPT and Lorentz symmetry with the neutral kaons}
Test of quantum mechanics (QM) can be performed by studying the time evolution of the quantum correlated  $\ks\kl$ system, in particular studying the interference pattern of the decay $\kl\ks \to \Ppip\Ppim\Ppip\Ppim$. The distribution of the difference decay times is given by:
\begin{equation}
I(|\Delta t|)\propto e^{-|\Delta t| \Gamma_L} +e^{-|\Delta t| \Gamma_S} -2cos(\Delta m |\Delta t|) e^{-\frac{\Gamma_s+\Gamma_L}{2}|\Delta t| }
\label{eq:deltat}
\end{equation}
One of the most direct ways to search for deviations from  QM
is to introduce a decoherence parameter $\zeta$ \cite{eberhard},
\ie\  multiplying by a factor $(1-\zeta)$ the interference
term in the last equation.
The definition of $\zeta$ depends on the basis chosen for the initial state~\cite{bertlmann1}
$|i \rangle  \propto  |K_S(+\vec{p})\rangle |K_L (-\vec{p})\rangle - 
 |K_L (+\vec{p})\rangle |K_S (-\vec{p})\rangle$ or $ |i \rangle   \propto  |K^0(+\vec{p})\rangle |\bar{K}^0 (-\vec{p})\rangle - 
 |\bar{K}^0 (+\vec{p})\rangle |K^0 (-\vec{p})\rangle. $

The case $\zeta=1$ ({\it i.e.} total decoherence) corresponds  
to the   spontaneous factorization of states
(known as Furry's hypothesis \cite{furry}).
Selecting a pure sample of  $\kl\ks \to \Ppip\Ppim\Ppip\Ppim$ and fitting eq.~\ref{eq:deltat} to data, KLOE has obtained  the following preliminary result based on 1$fb^{-1}$:
$\zeta_{SL}=  0.009 \pm 0.022_{\mbox{stat}}$ and $\zeta_{00}= \left( 0.03 \pm 0.12_{\mbox{stat}}\right)\times 10^{-5}$ consistent with QM predictions.

In a quantum gravity framework, space-time fluctuations
at the Planck scale ($\sim 10^{-33}\hbox{~cm}$), might induce a pure state to 
evolve into a mixed one \cite{hawk2}. This decoherence, in turn, necessarily
implies $CPT$ violation~\cite{wald}. 
In this context the CPT operator may be ``ill-defined'' and CPT   
violation effects might also induce a breakdown of the correlation in the initial state~\cite{mavro1,mavro2} which 
can be parametrized in general as:
$|i \rangle  \propto  |K_S(+\vec{p})\rangle |K_L (-\vec{p})\rangle - 
 |K_L (+\vec{p})\rangle |K_S (-\vec{p})\rangle \nonumber \\
 + \omega \left( |K_S (+\vec{p})\rangle |K_S (-\vec{p})\rangle - 
 |K_L (+\vec{p})\rangle |K_L (-\vec{p})\rangle\right)$
where $\omega$ is a complex parameter describing $CPT$ violation.
 Its order of magnitude might be at most  $|\omega| \sim \sqrt { (M^2_K/M_{Planck})/\Delta \Gamma } \sim 10^{-3}$, with $\Delta \Gamma = \Gamma_S - \Gamma_L$.
 KLOE has improved its  limit on  the $\omega$ parameter using about 1$fb^{-1}$.  The preliminary results, obtained  by fitting   the  $I(\Delta t;\Ppip\Ppim\Ppip\Ppim)$ distribution, are  
$\Re\omega =\left( -2.5^{+3.1}_{-2.3}\right)\times{10^{-4}}$ and
$\Im\omega =\left( -2.2^{+3.4}_{-3.1.} \right)\times{10^{-4}}$, consistent with quantum coherence  and $CPT$ symmetry. The accuracy reaches the interesting region of the Planck's scale.
\begin{figure}[h]
\begin{center}
\includegraphics[width=7.0cm]{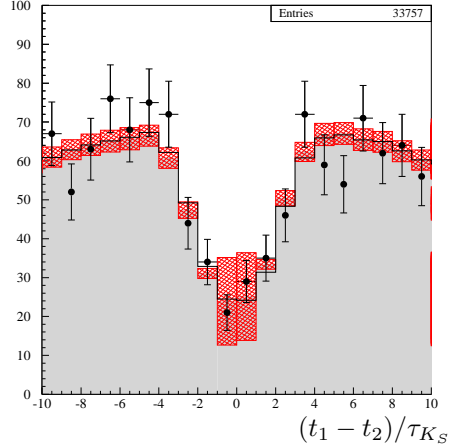}
\put(-80,10){ \small  $\boldmath (t_1-t_2)/\tau_{\ks}$}
\caption{Fit  of the difference $t_1-t_2$ of the decay times of $\ks\to \Ppip\Ppim$ and $\kl\to \Ppip\Ppim$, where $t_1$ is the time of the kaon having $\cos \theta>0$,  in the range $0<t_{\rm sid}<4$h. The black points  are the experimental data, the histogram is the fit results and the hatched area is the uncertainty arising from the efficiency, the resolution and the background evaluation  }\label{fig:fitlorentz}

\end{center}
\end{figure}

Another possibility for $CPT$ violation is based on spontaneous breaking of Lorentz symmetry in the context of the Standard Model Extension (SME)\cite{kostelecky1,kostelecky2}. In the SME $CPT$ violation  manifests to lowest order only in the  $\delta$ parameter, describing $CPT$ violation in the time evolution, which exhibits a dependence on the  kaon 4-momentum:
\begin{eqnarray}
\delta(p,\theta,t_{\rm sid}) &=&  
\frac{1}{2\pi} \int_0^{2\pi} \delta(\vec{p},t_{\rm sid})d\phi=\frac{i sin\phi_{SW} e^{i\phi_{SW}}\gamma}{\Delta m}  \\  &&\left (\Delta a_0  +\beta\Delta a_Z \cos\chi\cos\theta+ \beta\Delta a_Y \sin\chi\cos\theta\sin\Omega t_{\rm sid}+\beta\Delta a_X \sin\chi\cos\theta\cos\Omega t_{\rm sid}  \right )\nonumber
\end{eqnarray}
 after integration on $\phi$, where $\theta$ and $\phi$ are the conventional polar and azimuthal angles defined in the laboratory frame around the $z$ axis. $\Delta a_{\mu}$ are four $CPT$ and Lorentz symmetry violating coefficients for the two valence quarks, $\beta$ is the kaon velocity, $\gamma = 1/\sqrt{1-\beta^2}$, $\phi_{SW}$ is the superweak angle, $\chi$ is the angle between the $z$ laboratory axis and the Earth's rotation axis and  $\Omega$ is Earth's sidereal frequency. The sidereal time ($t_{\rm sid}$) dependence arises from the rotation of the Earth. KLOE has measured the $\Delta a_{X,Y,Z}$ parameters
by using the channel $\ks\kl\to \Ppip\Ppim\Ppip\Ppim$ and
 performing  an analysis on  the polar angle $\theta$ and the sidereal time $t_{sid}$. Fitting the distribution of the decay times difference   $I\left (t_1-t_2;\Ppip\Ppim (\cos \theta_1>0) \,\Ppip\Ppim (\cos \theta_2<0);t_{sid}\right)$  we obtain the preliminary results based on 1$fb^{-1}$: $\Delta a_X = (-6.3 \pm 6.0)\times 10^{-18} \, \mbox{GeV} $, $\Delta a_Y= (-2.8 \pm 5.9)\times 10^{-18} \mbox{GeV} $ and $\Delta a_Z= (-2.4 \pm 9.7)\times 10^{-18}\,\mbox{GeV} $. The result of the fit is shown  in fig.~\ref{fig:fitlorentz}.
A limit on the $\Delta a_0$ parameter has been obtained through the difference on the  \ks\ and \kl\  semileptonic charge asymmetry integrated on $t_{sid}$ and on  a symmetrical  polar angle region. Our preliminary result is $\Delta a_0 = (0.4 \pm 1.8)\times 10^{-17} \, \mbox{GeV}$

\section*{References}
\newcommand{\posk}[1]{PoS(KAON)#1}

\end{document}
